\documentclass[preprint, authoryear]{elsarticle}
\usepackage{amssymb}
\usepackage{graphicx}
\usepackage{dcolumn}
\usepackage{bm}

\journal{}

\begin{document}

\title{ Half metallic state and magnetic properties versus the lattice constant in  $Ti_{2}CoSn$ Heusler compound: an ab initio study.}
\author[label1]{A. Birsan}
\author[label1]{P. Palade}
\author[label1]{V. Kuncser}
\address[label1]{National Institute of Materials Physics, PO Box MG-07, Bucharest, Romania}

\begin{abstract}
The half metallic properties of $Ti_{2}CoSn$ full-Heusler compound is studied within the framework of the density functional theory with the Perdew Burke Ernzerhof generalized gradient approximation (GGA). Structural optimization was performed and the calculated equilibrium lattice constant  is 6.340 $\dot{A}$. The spin up band of compound has metallic character and  spin down band is semiconducting with  an indirect gap of 0.598 eV at equilibrium lattice constant. For the lattice parameter, ranging from 6.193 to 6.884 $\dot{A}$ the compound presents $100\%$ spin polarization and a total magnetic moment of 3$\mu_{B}$.  
\end{abstract}


\maketitle

\section{Introduction}
Discovered in 1903, when Fritz Heusler found that the composition $Cu_{2}MnAl$ behaves like a ferromagnet \cite{Heusler1903A, Heusler1903B}, Heusler alloys are half-metallic ferromagnets, semiconducting for electrons of one spin orientation and metallic character for electrons with the opposite spin orientation. 
These materials attracted remarkable attention because of the magnetic behavior or multifunctional magnetic properties, like magneto-optical \cite{vanEngen1983}, magneto-structural characteristics \cite{Kainuma2006} and nearly fully spin polarized conduction electrons, suitable for spintronic applications. Divided in two distinct families these ternary semiconducting or metallic materials have the general formula $XYZ$ or $X_{2}YZ$ with 1:1:1 (half-Heusler, ) or 2:1:1 (full-Heusler ) stoichiometry. 
The crystalline structure typical for half-Heusler materials is the non-centrosymmetric cubic structure $C1_{b}$ \cite{Heusler1903A,Heusler1903B,Webster1988}, while full-Heusler alloys crystallize in the cubic space group with $Cu_{2}MnAl$ ($L2_{1}$) prototype \cite{Heusler1903A,Heusler1903B,Webster1988} or $Hg_{2}CuTi$ prototype, if the number of 3d electrons of $Y$ atom exceeds  that of $X_{2}$ atom \cite{Kandpal2007}.    
In the last decades, $Mn_{2}$, $Co_{2}$, $Fe_{2}$ or $Cr_{2}$ -based Heusler alloys have been widely studied \cite{Buschow1981,Chen2006,Kumar2010,Graf2011,Weht1999,Ozdogan2006,Fecher2006,Galanakis2007,ZhongyuYao2010,Luo2007,Li2009}, however only few of $Ti_{2}$ -based full Heusler alloys were investigated \cite{KervanN2012,Pugaczowa-MichalskaMaria,KervanN2011,Lei2011}. In the present paper, the half metallic properties of $Ti_{2}CoSn$ are  theoretically investigation by first-principles calculations of density of states, energy bands, and magnetic moments. The purpose is to analyze if the half-metallicity reported in $Ti_{2}CoB$ \cite{KervanS2011},     $Ti_{2}CoGe$ \cite{Huang2012} and $Ti2CoZ$(Z=Al, Ga, In)\cite{Bayar2011,KervanN2011JPCS,KervanN2012JMMM,Wei2012,Zheng2012} is destroyed by the substitution of $B$, $Ge$, $Al$, $Ga$ or $In$ with $Sn$ atoms.   

\section{METHOD OF CALCULATION}
Geometric optimization and the electronic structure calculations, based on the density functional theory (DFT) were performed by means the self consistent full potential linearized augmented plane wave (FPLAPW) method implemented in Wien2k code \cite{WIEN2k}.  For the exchange and correlation interaction was used the Perdew Burke Ernzerhof \cite{Perdew} generalized gradient approximation (GGA) where the interstitial region employed Fourier series and for muffin-tin sphere, spherical harmonic functions. The plane wave cut-off value used was $K_{max}\*R_{MT} = 7$, where $K_{max}$ is the maximum modulus for the reciprocal lattice vector. The muffin-tin (MT) sphere radii selected for Ti, Co and Sn were 2.33, 2.33 and 2.35 a.u. respectively. The integration over the irreductible part of the Brilloin zone (BZ), using the energy eigenvalues and eigenvectors of a grid containing 2925k points, was done using the modified tetrahedron method \cite{Blochl}. The selected convergence criteria during self-consistency cycles considered was an integrated charge difference between two successive iterations less than 0.0001e/a.u.3 and the total energy deviation better than 0.01mRy per cell.

\section{RESULTS  AND DISCUSSIONS}
\begin{figure}
 \begin{center}
     \includegraphics[scale=0.4]{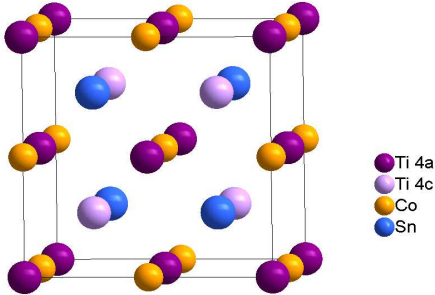}
  \end{center}
    \caption{The unit cell structure of $Ti_{2}CoSn$ compound has $Hg_{2}CuTi$ – prototype.  }
        \label{fig:Ti2CoSn}  
\end{figure}

\begin{figure}
 \begin{center}
     \includegraphics[scale=0.6]{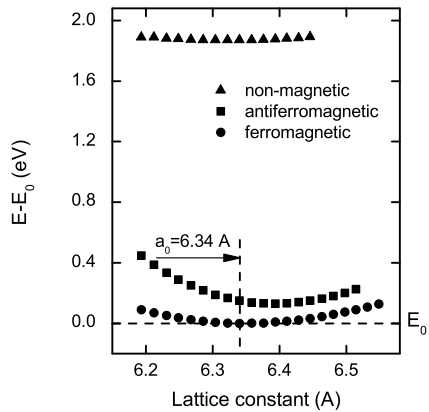}
  \end{center}
    \caption{Structural optimization for $Ti_{2}CoSn$. - The change of calculated total energyas function of the lattice constant is displayed The non-magnetic, antiferomagnetic and spin plarized ferromagnetic configurations are compared. For better comparison, alll energy scales are shifted to E(0)=0 eV.  }
        \label{fig:optimizareTi2CoSn}  
\end{figure}

The structure of  $Hg_{2}CuTi$ – prototype was used to define the unit cell of  $Ti_{2}CoSn$ compound, because Ti element is less electronegative than Co element.  In this structure, exhibited in Fig.\ref{fig:Ti2CoSn}, the Wyckoff positions $4a$, (0 0 0) and $4c$, (1/4 1/4 1/4) are occupied by Ti element, which makes them nearest neighbors, whereas  Co and Sn occupy the $4b$, (1/2 1/2 1/2)  and $4d$, (3/4 3/4 3/4) positions. The inversion symmetry typical for $L2_{1}$ structure with $Fm\bar{3}m$ is broken, and it is assigned to $F\bar{4}3m$ space group. Therefore, the $Hg_{2}CuTi$ – prototype, also called inverse Heusler structure can be thought of as a generalization of $L2_{1}$ , which is the typical Heusler structure.  The GGA parameterization with non-magnetic, antiferromagnetic and spin polarized ferromagnetic setup were employed for structural optimization, within the FLAPW-scheme. The superstructure constructed to be used for these configurations, resulted from spreading the unit cell crystal structure along the axis of $a$ lattice parameter and contains an even number of $Co$ atoms with nearest-neighbor spins, oriented in opposite directions. Geometrical optimization, plotted in Fig.\ref{fig:optimizareTi2CoSn} was performed to determine the lattice parameter which minimizes the total energy. 
 The iteration procedure for each of the different lattice parameter was performed up to self-consistency.  In the non-magnetic  and antiferromagnetic cases, the minimum total energies were found to be higher  than that found for ferromagnetic (spin-polarized) configuration.  The optimized lattice constant found for $Ti_{2}CoSn$ compound is 6.34 $\dot{A}$, in ferromagnetic phase and no experimental or theoretical value are available in literature, to compare with our results. 

\begin{figure}
 \begin{center}
         \includegraphics[scale=0.7]{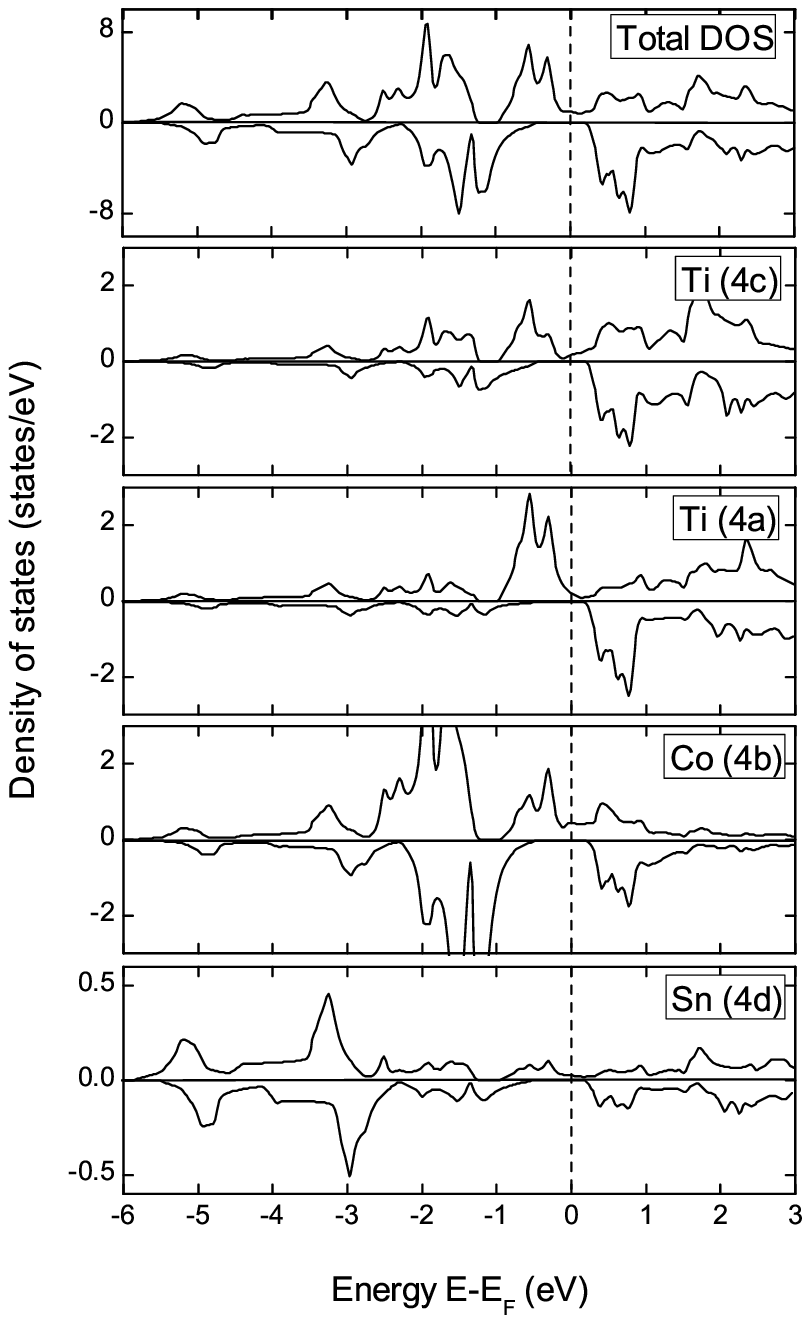}  
    \end{center}
    \caption{Spin-projected total densities of states (DOS) and partial DOSs calculated at predicted equilibrium lattice constant.}
       \label{fig:totaldosTi2CoSn}
\end{figure}

The spin-projected total densities of states (DOS) and partial DOSs(PDOS) of  $Ti_{2}CoSn$ obtained by performing the spin-polarized calculations at equilibrium lattice constant are displayed in Fig. \ref{fig:totaldosTi2CoSn}.  The majority spin channel (spin-up band) presents a clear metallic character while the minority spin channel (spin-down band) is semiconducting with a energy gap around the Fermi level. Additionally are exhibited the main partial densities of states of Ti, Co and Sn atoms, where 4c, 4a, 4b and 4d represent the Wyckoff positions from crystal structure. It is worth to be noted, that between -2 eV and -1 eV the energy  comes mainly from d electrons of Co atoms. Between -5.5 eV and -2.5 eV the significant contribution comes from  p electrons of Sn. 

\begin{figure}
 \begin{center}
   \includegraphics[scale=1.0]{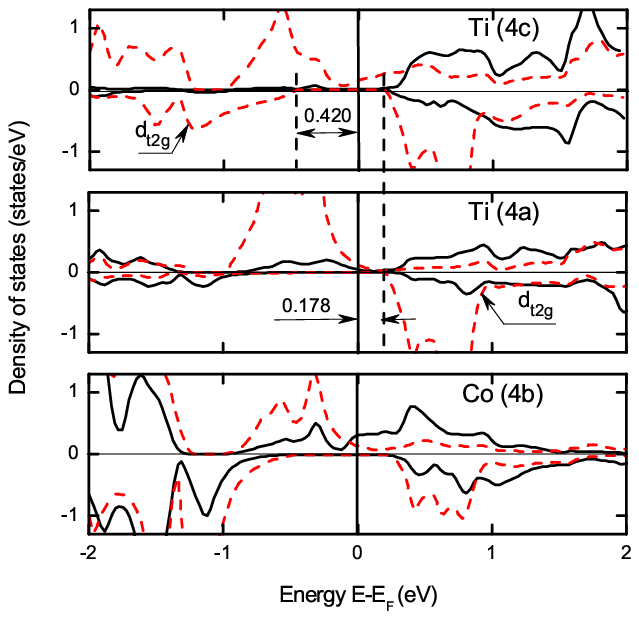}  
 \end{center}
    \caption{The main partial densities of states at optimized lattice parameter of $Ti_{2}CoSn$, $d_{eg}$ and $d_{t2g}$ being indicated by solid and dashed line, respectively}
    \label{fig:sitedosTi2CoSn}
\end{figure}

At equilibrium lattice constant the Fermi level is located at 0.393 eV from the triple degenerated occupied Ti(4c) $d_{t2g}$ (below $E_{F}$) and at 0.151 eV from unoccupied Ti(4a) and Ti (4c) $d_{t2g}$  states (above $E_{F}$) as has been plotted in Fig \ref{fig:sitedosTi2CoSn}. Therefore, the energy gap is clearly formed due to Ti-Ti hybridization, by splitting d states of Ti atoms, located in the two different Wyckoff positions in the minority spin channel. The $3d$ orbitals of Co are fully occupied and form a weak covalent interaction with Ti $3d$ orbitals. 

\begin{figure}
 \begin{center}
    \includegraphics[scale=0.8]{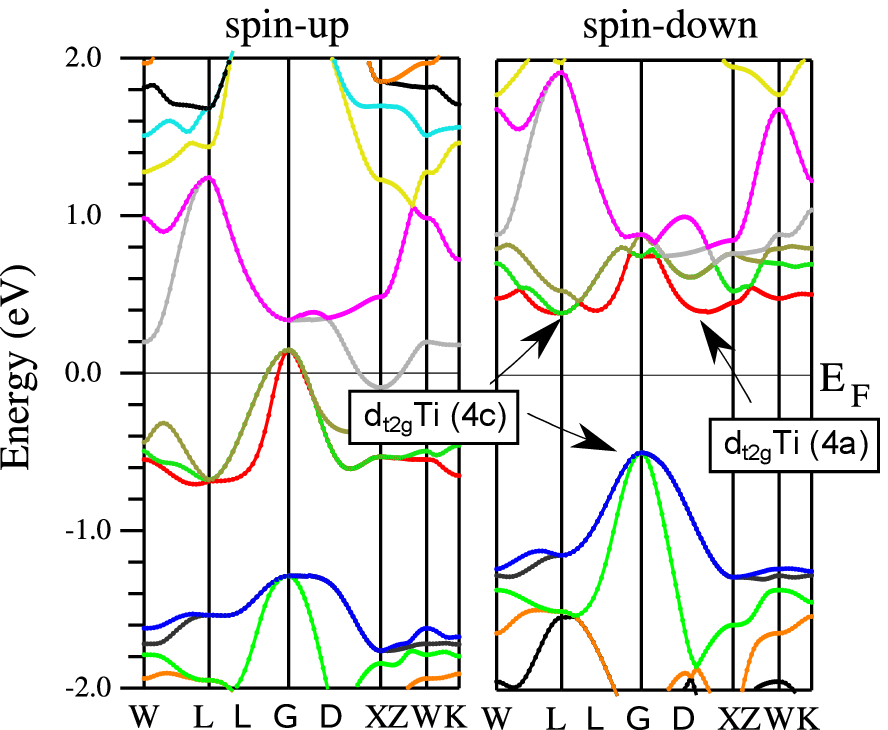} 
      \end{center}
   \caption{The band structure of $Ti_{2}CoSn$ for spin-up (left panel) and spin-down (right panel) electrons for geometrical optimized structure}
    \label{fig:bandTi2CoSn}
\end{figure}

In Fig \ref{fig:bandTi2CoSn} is plotted the band structure of $Ti_{2}CoSn$ compound at equilibrium geometry. The width of the indirect band gap of 0.598 eV is calculated using the energy of the highest occupied band at the G point (0.420 eV) and the lowest unoccupied band at L point (0.178 eV), from the spin-down band structure. The bandstructure reveals that Ti-Ti bonding interactions are the strongest, even though Ti-Co interactions also play a role. The size of the gap, related to the difference in energy, between the bonding and antibonding $d$ states it is larger than the reported energy gap for all $Ti_{2}Ni$, $Ti_{2}Fe$, $Ti_{2}Mn$-based compounds studied \cite{Lei2011,KervanN2011JPCS,Wei2012,Zheng2012}. The substitution of $B$, $In$ or $Ge$ with $Sn$ in $Ti_{2}Co$-based compounds, leads to a decrease of minority band gap \cite{KervanN2012,KervanS2011,Huang2012} at reported optimized lattice constants but doesn't destroy the half metallic character.  For  $Ti_{2}CoAl$ compound, the size of energy gap from minority spin channel reported by \cite{Wei2012} (0.68 eV) is significantly larger than those calculated by other authors: 0.49 eV \cite{Bayar2011} and 0.486 eV \cite{Zheng2012}. A similar situation was found for $Ti_{2}CoGa$ when the sizes of the minority band gaps were 0.5eV \cite{KervanN2012JMMM} and  0.68 eV \cite{Wei2012}. In this context, further investigations are needed to clarify the width of band gaps from minority spin channel in the $Ti_{2}CoZ$ compounds with Z=$Al$ or $Ga$.    

\begin{figure}
 \begin{center}
    \includegraphics[scale=0.8]{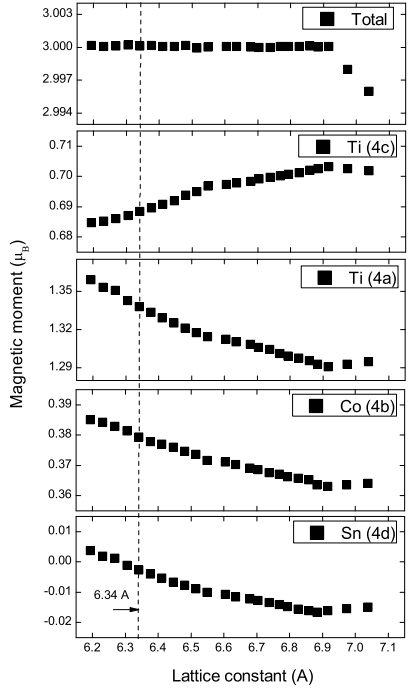}
      \end{center}
   \caption{The total and site-specific magnetic moments of $Ti_{2}CoSn$ compound as function of lattice constant.}
    \label{fig:magneticmomentTi2CoSn}
\end{figure}

Fig. \ref{fig:magneticmomentTi2CoSn}  shows the calculated values for total spin magnetic moment $M^{calc}_{tot}$ ($\mu_{B}/f.u.$) and spin magnetic moment of each atom $M^{calc}$ ($\mu_{B}/atom$) as function of lattice parameter. The total magnetic moment is equal to 3 $\mu_{B}$, following the  rule $M_{t} = Z_{t}-18$ $\mu_{B}/f.u.$ \cite{Zheng2012} ($M_{t}$  is the total spin magnetic moments per formula unit cell and $Z_{t}$, the total number of valence electrons) and starts decreasing for a lattice constant above 6.884 $\dot{A}$. 
In the lattice parameter interval where the total spin magnetic moment is constant, the spin magnetic moments of Ti (4a), Co and Sn decrease with increasing of  lattice constant, while the magnetic moment of Ti (4c) increases. The spin-polarization calculations reveal that the site-resolved magnetic moments per atom, at the optimized lattice constant are 0.918,   1.323, 0.177 and  -0.0006 $\mu_{B}$ for Ti(4c), Ti(4a), Co and Sn, respectively.  A considerable contribution to the total magnetic moment comes from the interstitial region (0.582 $\mu_{B}$). 

\begin{figure}
 \begin{center}
    \includegraphics[scale=0.65]{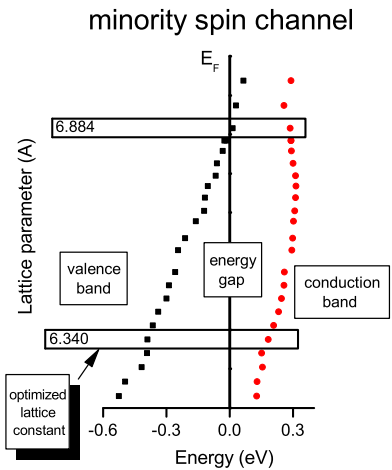}
      \end{center}
   \caption{The positions of lowest unoccupied states from the conduction band (solid red circles) and of highest occupied states from valence band (solid black squares) of total DOSs from minority spin channel, for $Ti_{2}CoSn$ as function of lattice parameter.}
    \label{fig:gapTi2CoSn}
\end{figure}

The size of the band gaps as function of lattice parameter is displayed in Fig \ref{fig:gapTi2CoSn}. The width of it decreases with the increase of lattice parameter. Furthermore, the more relaxed the structure become the Fermi level is shifted from the bottom edge of the minority conduction band to the upper edge of the minority valence band,  which leads to a half metallic character  for a lattice parameter up to 6.884 $\dot{A}$, when this property is lost and the $Ti_{2}CoSn$ compound presents not only in majority but also in minority channel a clear metallic character.

In this context, the presence of energy gap in the minority spin channel leads to a $100\%$ spin polarization for bulk $Ti_{2}CoSn$. The polarization remains constant for the lattice parameter interval where the compound has half metallic property, which makes the $Ti_{2}CoSn$ an interesting material for future application in magnetoelectronics and spintronics.

\section{CONCLUSIONS}
The electronic band structure and magnetic properties of $Ti_{2}CoSn$ were analyzed with the self-consistent full-potential linearized augmented-plane-wave basis scheme based on the density functional theory with the generalized gradient approximation (GGA). The results of spin-polarized calculations present the half-metallic ferromagnetic nature of $Ti_{2}CoSn$ with a indirect band gap in the minority spin channel (0.589 eV) at  calculated equilibrium lattice constant (6.340 $\dot{A}$). The magnetic moment of $Ti_{2}CoSn$ material is 3 $\mu_{B}$, fully polarized in a lattice parameter range of 6.193 $\div$ 6.884 $\dot{A}$.

\section{ACKNOWLEDGMENTS}
This work was financially supported from the projects  PNII IDEI 75/2011 and Core Program PN09-450103 of the Romanian Ministry of Education Research, Youth and Sport. 

\bibliographystyle{elsarticle-harv}

\end{document}